\documentclass[sigconf]{acmart}
\AtBeginDocument{%
  }

\copyrightyear{2025}
\acmYear{2025}
\setcopyright{rightsretained}
\acmConference[CHI EA '25]{Extended Abstracts of the CHI Conference on Human Factors in Computing Systems}{April 26-May 1, 2025}{Yokohama, Japan}
\acmBooktitle{Extended Abstracts of the CHI Conference on Human Factors in Computing Systems (CHI EA '25), April 26-May 1, 2025, Yokohama, Japan}\acmDOI{10.1145/3706599.3719871}
\acmISBN{979-8-4007-1395-8/2025/04}

\begin{document}

\title{Bridging Bond Beyond Life: Designing VR Memorial Space with Stakeholder Collaboration via Research through Design}


\author{Heejae Bae}
\email{hj_bae@kaist.ac.kr}
\orcid{0009-0009-7305-8899}
\affiliation{%
  \institution{Department of Industrial Design, KAIST}
  \city{Daejeon}
  \country{Republic of Korea}
}

\authornote{Both authors contributed equally to this research.}

\author{Nayeong Kim}
\authornotemark[1]
\email{nayeong_k@kaist.ac.kr}
\orcid{0000-0002-3573-5547}
\affiliation{%
  \institution{Department of Industrial Design, KAIST}
  \city{Daejeon}
  \country{Republic of Korea}
}

\author{Sehee Lee}
\email{leesh9457@kaist.ac.kr}
\orcid{0009-0006-2661-8513}
\affiliation{%
  \institution{Department of Industrial Design, KAIST}
  \city{Daejeon}
  \country{Republic of Korea}
}

\author{Tak Yeon Lee}
\email{takyeonlee@kaist.ac.kr}
\orcid{0000-0002-9235-9947}
\affiliation{%
  \institution{Department of Industrial Design, KAIST}
  \city{Daejeon}
  \country{Republic of Korea}
}

\begin{abstract}
    The integration of digital technologies into memorialization practices offers opportunities to transcend physical and temporal limitations. However, designing personalized memorial spaces that address the diverse needs of the dying and the bereaved remains underexplored. Using a Research through Design (RtD) approach, we conducted a three-phase study: participatory design, VR memorial space development, and user testing. This study highlights three key aspects: 1) the value of VR memorial spaces as bonding mediums, 2) the role of a design process that engages users through co-design, development, and user testing in addressing the needs of the dying and the bereaved, and 3) design elements that enhance the VR memorial experience. This research lays the foundation for personalized VR memorialization practices, providing insights into how technology can enrich remembrance and relational experiences.
\end{abstract}





\maketitle

\section{INTRODUCTION and BACKGROUND}

\begin{quote} 
\textit{"When there’s no one left in the living world who remembers you, you disappear from this world. We call it the final death."} - From the Disney film \textit{Coco} \cite{Coco2017} 
\end{quote}
Remembering and being remembered have always been universal human needs. From the Great Pyramids of ancient Egypt to the 9/11 Memorial, humans have long constructed physical monuments to collectively mourn, honor, and share the experience of death and loss \cite{3Moncur2014}. In the digital age, the integration of technology into memorialization has emerged as a pivotal research area in HCI. Early work in this field introduced concepts such as “techno-spiritual practices” \cite{4Bell2006, 5Wyche2006} and “thanatosensitive design” \cite{6Massimi2009}, which emphasize designing systems sensitive to mortality, dying, and remembrance. Inspired by these pioneers, many academic and industry projects have explored how digital systems might help shape the way people look back and interpret people’s lives after they have passed away \cite{1facebook,7Gulotta2017,8Kwon2021}. 

While online memorials provide spaces to share pictures and stories of the deceased, they remain static and non-immersive, limiting deep engagement \cite{39Lopes2014}. Similarly, CemTech has explored enhancing cemetery experiences with digital tools, yet these innovations are often site-specific and limited to augmenting existing physical spaces \cite{40Allison2024}. In contrast, VR uniquely enables users to step into an immersive, interactive space, offering a sense of presence and engagement that traditional digital memorials or CemTech applications cannot fully capture. 

In this context, Virtual Reality (VR) technology has been explored for its potential for commemorative purposes. For example, museums and memorial sites have leveraged 3D digital projections or VR technology to offer visitors deeply engaging public memorial experiences \cite{11Hakkila2019,12Hakkila2024}. Additionally, concepts like SenseVase \cite{13Uriu2021} have combined physical rituals with virtual tributes to explore how embodied rituals can enrich virtual memorialization. However, much of the studies focus on ‘online public commemoration,’ and despite these advancements, little attention has been given to leveraging VR technology to reflect the unique memories and emotions of an individual.

To our knowledge, the only notable example of using VR for personal memorialization is a South Korean TV documentary\footnote{I Met You: Mother meets her deceased daughter through VR technology (https://www.youtube.com/watch?v=uflTK8c4w0c\&t=1s).} where a grieving mother interacted with a VR recreation of her deceased daughter. However, this example was more of a media experiment than a design study, leaving academic discussions on how to design and evaluate personalized VR memorial spaces still in the early stages.

A memorial space should go beyond the place for remembering the deceased; it should harmonize the identity of the dying with the emotional needs of the bereaved \cite{20Petersson2011,21Maddrell2013}. The dying reflect on their life, organize their digital legacy, and consider how they wish to be remembered \cite{22Albers2023}, while the bereaved seek to maintain bonds with the deceased and revisit memories to sustain an emotional connection \cite{25Kim2024}. However, thanatosensitive designs often prioritize one group over the other. For instance, studies on digital legacy emphasize preserving and organizing personal data \cite{17Gulotta2013,18Locasto2011,19Gulotta2017} but have limited attention to how the dying might interpret and utilize these legacies. Meanwhile, grief-support designs for the bereaved tend to focus on the reinterpretation of the dying’s memory after their passing \cite{25Kim2024, 27Wendy2015, 36She2021}, failing to fully capture their intentions or identity.

Death inherently separates the dying and the bereaved across time, but meaningful memorialization requires a structure that bridges this divide. Janet et al. highlighted that death-related experiences, such as digital legacy management, are not solely individual matters but are enriched through collaborative processes with family, friends and acquaintances \cite{16Janet2021}, suggesting the importance of collaboration. Therefore, a collaborative process is needed to integrate the different needs of both the dying and the bereaved. Our study aims to explore the way to incorporate the complex needs, thereby enhancing the meaning and value of memorial experiences. To integrate the complex needs of both parties, the process should take place while both are still alive. Therefore, in this study, we have identified two stakeholders: the Owner, who designs the space from the perspective of the dying, and the Visitor, who represents the bereaved, provides feedback and collaboratively shapes the identity of the space.

To this end, we adopt a Research through Design (RtD) \cite{14Zimmerman2007, 15Gaver2012} approach, designing and evaluating a personalized VR memorial space with two stakeholders, leveraging VR’s spatial, immersive, and interactive qualities to explore how they can enhance memorial experiences beyond conventional digital formats. RtD is particularly suited to this study as it facilitates the inquiry of underexplored domains, enabling initial exploration of VR memorialization as a novel design space. By treating VR memorial space as a central artifact, RtD allows us to explore the emotional and relational dimensions of memorialization practices.
Specifically, our study addresses the following three research questions:
RQ1) What value does a VR memorial space, created through stakeholder participation, provide?
RQ2) What should be considered in the process of designing a VR memorial space that integrates stakeholders?
RQ3) What design elements should be considered to create a VR memorial space that integrates stakeholders?
To achieve this, our study followed a three-phase process: \textbf{Participatory Design}, where users actively engaged in envisioning memorial spaces; \textbf{Development}, focused on implementing a VR memorial space; and \textbf{User Testing}, where eight participants provided experiential evaluations.

The contributions of this study are twofold: 1) Expanding academic discourse by proposing and validating a novel design process for personalized VR memorial spaces, addressing the gaps in current thanatosensitive design research. 2) Demonstrating how remembrance can transcend physical spaces and extend into digital and virtual spaces, offering new possibilities for the development of memorial cultures and the creation of novel experiences of memorialization.

\section{METHODOLOGY}

\subsection{Procedure Overview}
This study adopted a Research through Design (RtD) approach, structured into three phases, to uncover key design implications~— including features, processes, and decision-making strategies~— necessary The research process centered on two primary stakeholders: the Owner and the Visitor. The Owner is a living individual who designs the identity and core contents of his/her own memorial space from the perspective of the dying. The Visitor participates in the design process from the perspective of the bereaved by previewing the VR space in its early stages, providing feedback on the potential content, and shaping the space’s identity together. As illustrated in Figure 1, the process begins with Participatory Design (phase 1), which includes reflective activities to capture the Owner’s requirements and expectations for her own memorial space (via Memory Kit), as well as collaborative design sessions where the Owner and the Visitor share perspectives and negotiate the design of the VR space. The process continues to the development of the VR memorial space (phase 2) and concludes with its exploratory evaluation (phase 3), where additional Visitors explore the space and provide feedback.

\begin{figure*}[tbp]
    \centering
    \includegraphics[width=1\linewidth]{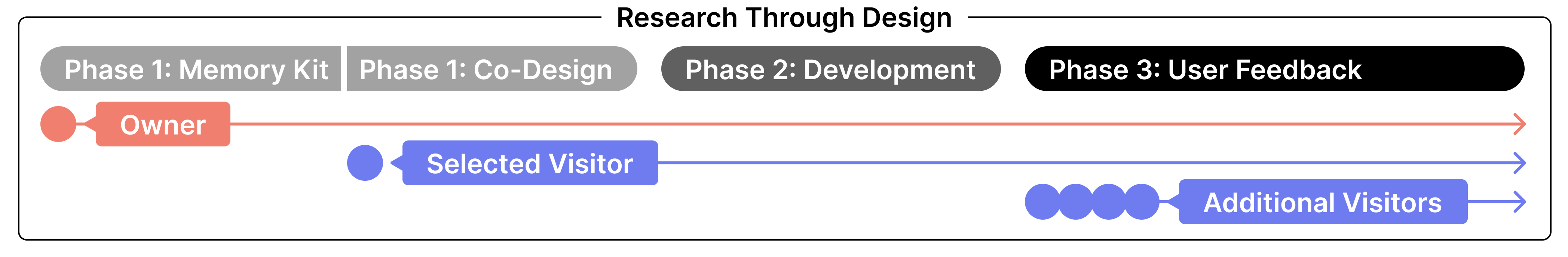}
    \caption{An overview of the three-phase RtD process for designing a VR memorial space  with stakeholder collaboration}
    \label{fig:process}
    \Description{A timeline illustrating the three phases of the Research through Design process: Phase 1 includes “Memory Kit” and “Co-Design,” where the Owner and Selected Visitor participate. Phase 2 focuses on “Development,” involving ongoing contributions from the Owner and Selected Visitor. Phase 3 is “User Feedback,” where Additional Visitors join, overlapping with the contributions of the Owner and Selected Visitor.}
\end{figure*}

\subsection{Participant}

\begin{table*}
  \caption{Participant demographics, roles, and phase-specific participation}
  \label{tab:participant_info}
  \begin{tabular}{cccccccc}
    \toprule
   Group&ID&Gender&Age&Role&Memory Kit&Co-Design&User Testing\\
    \midrule
1 & P1O1 (Owner)  & Female & 56 & Owner              & O & O & X      \\
1 & P2V1 (Daughter)  & Female & 26 & Visitor           & X & O & X   \\
2 & P3O2 (Owner)  & Female & 49  & Owner              & O & O & O       \\
2 & P4V2 (Daughter)  & Female & 25 & Visitor            & X & O & O \\ 
2 & P5V3 (Sister)  & Female & 54 & Visitor             & X & X & O \\ 
2 & P6V4 (Sister)  & Female & 51  & Visitor        & X & X & O \\ 
2 & P7V5 (Husband)  & Male   & 56 & Visitor           & X & X & O \\ 
2 & P8V6 (Brother-in-Law)  & Male & 55  & Visitor    & X & X & O  \\ 
  \bottomrule
\end{tabular}
\end{table*}

In phase 1, two Owners were initially recruited to conduct the \textit{Memory Kit}. After that, each Owner selected a trusted family member as their Visitor to participate in the \textit{Co-Design Session}. From phase 2 onward, only one group (Group 2) continued to participate due to participant availability. In phase 3, additional Visitors—family members of the Owner—participated in experiencing the VR memorial space. Participants were labeled using a combination of their participant number and role (e.g., P1O1 for Owner, P2V1 for Visitor). Table \ref{tab:participant_info} summarizes the demographics of the participants and their involvement across the research phases.

Group 1 consisted of a mother-daughter pair: P1O1 (Owner) and P2V1 (Visitor). Group 2 included P3O2, the Owner of the finalized VR memorial space, P4V2 and four additional family members (P5V3, P6V4, P7V5, and P8V6) as a Visitor. All participants were recruited through personal networks and provided informed consent prior to their involvement.

\section{Research through Design}

\subsection{Phase 1: Participatory Design}
The design of the VR memorial space began with a Participatory Design (PD) \cite{38Simonsen2012} approach, consisting of reflective engagement using the \textit{Memory Kit} and the \textit{Co-Design Session}. This approach ensured that the space reflected the Owners’ intentions while incorporating the perspectives and expectations of Visitors.

The first activity, \textit{Memory Kit}, was designed to help Owners reflect on their lives and revisit significant memories to select meaningful objects and content for their VR memorial space. Given that most Owners had not deeply contemplated their mortality or memorial planning \cite{30Albers2024}, a structured 7-day diary was provided to guide this reflective process and collect relevant data. The diary included three categories of prompts: 1) To help Owners deeply contemplate their mortality (e.g., “Where would you wish to visit before the end of your life?”, “What would you like to achieve before the end of your life?”), 2) To guide the Owners to conceptualize the memorial space (e.g., “What message would you like to convey to Visitors?”, “What atmosphere would you like Visitors to experience?”), and 3) To help them identify and select meaningful artifacts (e.g., “Are there any photos or videos you want to include in the memorial space?”, “What is the most valuable item you want to leave behind?”). Participants received a set of daily envelopes, each containing a question sheet and a response page. They were instructed to open one envelope per day on the designated date and record their answers. This process resulted in the collection of 85 photos, 1 video, and textual data derived from the kit responses of the two participating Owners. 

The second activity, \textit{Co-Design Session}, was a 120-minute workshop where the Owner and the Visitor collaborated to align their perspectives and co-design the VR space’s elements (e.g., color scheme, structure), content (e.g., mementos, messages), and interactions. Two groups were formed, with each Owner choosing a family member to participate as a Visitor. 
To maintain spontaneous communication between participants, the moderator’s role was intentionally limited, focusing on time management, task instructions, and facilitating discussions only when necessary. For the first 30 minutes, each Owner shared their vision for the memorial space by presenting photos collected from the Memory Kit and elaborating on the associated memories. Then, visitors articulated their requirements and expectations regarding the data and the spatial design (30 minutes). In the final 60 minutes, the participants discussed and determined the content to be uploaded to the space (e.g., objects, photos, and videos) and collaboratively sketched its elements, such as structure, composition, and interactions, thereby finalizing the design direction. This activity resulted in sketches of the VR memorial space along with audio recordings, capturing insights from both participants.

\subsection{Phase 2: Development}
\begin{figure*}[tbp]
    \centering
    \includegraphics[width=1\linewidth]{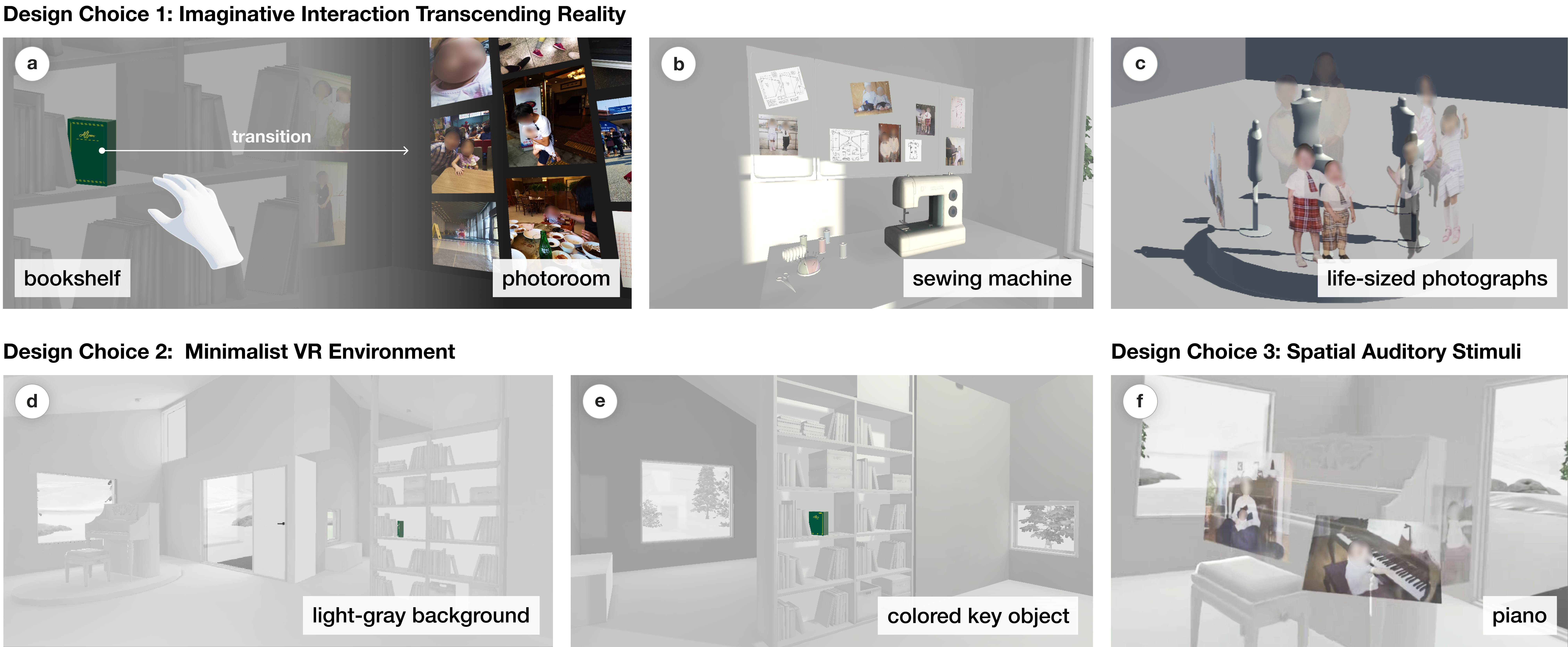}
    \caption{Design choices in the VR memorial space}
    \label{fig:design_choices}
    \Description{The figure illustrates three design choices for a VR memorial space. 1.	Design Choice 1: Imaginative Interaction Transcending Reality: (a) A transition from a bookshelf to a photoroom is shown, demonstrating how users can interact with virtual objects. (b) A sewing machine setup, surrounded by personal memorabilia such as photos and sewing patterns. (c) Life-sized photographs of individuals displayed in the virtual space, creating a sense of presence. 2. Design Choice 2: Minimalist VR Environment: (d) A VR room with a white background to create a simple and distraction-free space. (e) A bookshelf with a single colored key object highlighted amidst other muted objects. 3. Design Choice 3: Spatial Auditory Stimuli: (f) A piano in a room with large windows, where spatial auditory cues complement the displayed visual elements, such as photos.}
\end{figure*}

After finishing phase 1, we continued with the second group based on participants' availability. Using the data collected during phase 1, including physical materials (e.g., photos and videos), as well as requirements and expectations, we designed and implemented a VR memorial space to meet their needs. The design aimed to accurately capture and represent the Owner's identity, values, and memories. At the same time, we prioritized to offer an immersive experience for the Visitors, allowing them to deeply connect with the memorial space, by leveraging the unique affordances of VR technology. 
To achieve the design goals, we formulated three design choices outlined below, drawing from prior literature on VR space design while implementing the memorial space. Figure \ref{fig:design_choices} illustrates the three key design choices in the VR memorial space, with specific elements detailed in Fig. 2a through Fig. 2f.

\textit{\textbf{Design Choice 1: Imaginative Interaction Transcending Reality}} 
- VR environments are most compelling when they offer imaginative or even surreal experiences \cite{31Husung2019}. Based on this understanding, the VR memorial space included interactions that transcend reality \textbf{to explore whether these interactions could effectively convey the Owner’s identity and the personal meaning behind cherished objects and memories, while fostering emotional immersion and active exploration for the Visitor.} To achieve this, several interactive features were implemented. For instance, taking a book off the bookshelf initiated a transition to a new space (Fig. 2a), and approaching a sewing machine automatically triggered the creation of a curtain (Fig. 2b). And palm-sized photographs were enlarged to life-sized visuals, allowing Visitors to feel as though they were encountering the people in the photograph (Fig. 2c).

\textit{\textbf{Design Choice 2: Minimalist VR Environment}}
- Reducing visual complexity and emphasizing key elements can enhance user focus and engagement while mitigating distractions caused by excessive detail \cite{Guo2021EffectsOV}. In light of this finding, this design choice aimed \textbf{to explore how a minimalist approach influences the Visitor's immersion and engagement with key content.} To help Visitors focus on meaningful objects, we intentionally rendered most background elements, such as walls and floors, in light-gray (Fig. 2d). In contrast, the key objects were highlighted with distinct colors and textures to naturally guide the visitor's focus. For example, the single interactive album was vividly highlighted in green to draw attention and encourage exploration (Fig. 2e).

\textit{\textbf{Design choice 3: Spatial Auditory Stimuli}} - Spatial auditory elements are powerful tools for evoking emotional immersion and storytelling \cite{33Bhide2019}. Hence, this design choice aimed \textbf{to explore how spatial auditory stimuli could deepen the Visitor’s emotional engagement and facilitate the natural recall of shared memories with the Owner.} The VR memorial space incorporated auditory cues that responded to the Visitor’s proximity and interaction with specific objects. 
For example, as the Visitor approached the piano in the living room, Yuhki Kuramoto’s "Lake Louise", a piece often played by the Owner, began playing softly (Fig. 2f). The closer the Visitor moved to the piano, the louder and clearer the music became, transforming from a subtle background melody to a prominent and immersive focal point.

\subsection{Phase 3: User Testing}
Phase 3 involved the Owner and Visitors experiencing the initial implementation of the VR memorial space. This phase focused on examining whether the Owner’s intentions were effectively reflected in the space and how the design choices facilitated emotional connections and immersive experiences of the Visitors.
The participant group included the Owner (P3O2) and the original Visitor from phase 1 (P4V2), along with four additional family members selected by the Owner (P5V3, P6V4, P7V5, P8V6). Participants explored the VR space freely for approximately 20 minutes using a VR Head-Mounted Display (Meta Quest 3\footnote{https://www.meta.com/at/en/quest/quest-3/}). Following the exploration, semi-structured interviews, lasting about 30 minutes, were conducted individually to collect feedback on the space’s emotional, interactive, and usability aspects. The entire process for each participant was documented through VR screen recordings and video recordings for further analysis.

\section{VALUE OF VR MEMORIAL SPACE AS BONDING MEDIUM}
The design of the VR memorial space demonstrated its potential to go beyond being merely a commemorative space, functioning as a relational medium that facilitated meaningful connections between stakeholders. This section explores the key values created through this process, addressing \textbf{RQ1}: What value does a VR memorial space, created through stakeholder participation, provide? 

\paragraph{\textbf{Bridging connections among visitors}} 
The VR memorial space revealed its potential to sustain and even 
strengthen relationships between stakeholders after the Owner’s passing. Participants viewed that the space was not just a static repository of memories but a dynamic medium where content could be enriched through contributions and interactions among Visitors. For instance, P5V3 noted, \textit{"If other Visitors could add their memories to the space, the content would grow richer, and it might encourage me to visit more often."} This highlights that updating personal memories not only deepens the content but also fosters a sense of connection among Visitors, encouraging repeated visits.
According to P2V1, \textit{"Whether it’s about mother or just catching up, having conversations with others in that space would be meaningful. Through that, we could naturally know how everyone is doing after her passing."} This indicates that the space could facilitate conversations and provide a way for Visitors to stay updated on each other's lives after the Owner's passing.

\paragraph{\textbf{Facilitating a deeper understanding of the owner}}
The VR memorial space offered Visitors with an opportunity to discover previously unknown aspects of the Owner's life. 
P2V1 described the space as \textit{“continuing funeral hall for my mom,"} where meeting friends and family could naturally lead to conversations that \textit{"branch out into different stories.”} Similarly, P4V2 noted that visiting the space with others would likely \textit{“end up talking about my mom—sharing personal stories or episodes that each of us remembers.”} These interactions could provided Visitors with a richer and deeper understanding of the Owner’s multifaceted life.

\paragraph{\textbf{Fostering new bonds between stakeholders}}
The VR memorial space demonstrated its potential to create new forms of bonding between Visitors who had no direct relationship with the Owner (e.g., child’s spouse or grandchildren) and those who shared close connections with the Owner. 
For example, P2V1 noted, \textit{“It would be nice to visit with my future husband or family. Instead of explaining my memories myself, this space would let them see for themselves and understand, ‘Oh, so you have these kinds of memories with your mom.’”} This highlights that the VR memorial space serves not only as a platform for commemorating the Owner, but also as a medium for narrative exchange and relationship-building among Visitors, bridging generational or relational gaps.

\section{REFLECTION ON DESIGN PROCESS}

The design process of the VR memorial space provided meaningful insights into how stakeholder's participation influenced its creation. This section addresses \textbf{RQ2}: What should be considered in the process of designing a VR memorial space that integrates stakeholders? It reflects on how stakeholder participation shaped its preparation, collaboration, and relational balance. 

\paragraph{\textbf{Importance of preparing memorial space before death}} The process of preparing VR memorial space goes beyond simply collecting data, offering Owners an opportunity to reflect on and reinterpret their lives and identities. Selecting memories, objects, and photos for the space helped Owners revisit their past and gain a more positive perspective on life and death. For instance, P1O1 noted, \textit{“This process helped me stay calm and prepared, even if death were to come suddenly,”} while P3O2 mentioned, \textit{“It gave me a chance to recall forgotten memories and reflect on what truly matters to me.”}
To make this process even more meaningful, it is important to consider designing structured and carefully designed guides to support meaningful reflection and the organization of their memories.

\paragraph{\textbf{Designing memorial space through collaboration of the owner and the visitor}} The process of designing the VR memorial space revealed differences in priorities and perspectives between Owners and Visitors. Owners tended to focus on shared memories, while Visitors expected the space to reflect more of the Owner’s personal aspects. For instance, P3O2 selected lots of family photos to emphasize shared memories. In contrast, P2V1 questioned the inclusion of so many photos featuring her (Visitor), asking, \textit{“What is the point of seeing me when I visit to remember the Owner?”} Similarly, P4V2 expressed a desire to see more personal aspects of the Owner, saying, \textit{“I wish there were more personal photos, like ones from her (Owner’s) childhood.”}
This collaborative process was meaningful not only as a step in designing the space but also as an opportunity for Owners and Visitors to share and reconcile their perspectives while co-creating the narratives embedded in the space. Based on this finding, designing collaboration processes that support such exchanges is important to balancing diverse expectations and ensuring that the resulting space reflects both shared and personal memories in a cohesive way.

\paragraph{\textbf{Balancing relationships in VR memorial space}} The relationship between Owners and Visitors influenced the contents in the memorial space. Owners often prioritized memories with close family members or limited editing rights to specific Visitors. However, Visitors who considered themselves close to the Owner expressed feelings of exclusion or disappointment when they discovered that content related to them was missing. For instance, P5V3, after visiting the space, remarked, \textit{“She (Owner) only included her family photos. She should have added memories with me too—I am her sibling! Not seeing any photos of us made me feel a bit excluded.”}
When designing such spaces, it is important to consider that Visitors with diverse types of connections with the Owner can visit and to explore ways to make them feel included and deeply engaged in remembrance.

\section{REFLECTION OF DESIGN CHOICES}

The design choices implemented in the VR memorial space highlighted how specific design elements shape user experiences and interactions. This section addresses \textbf{RQ3}: What design elements should be considered to create a VR memorial space that integrates stakeholders?
It reflects on how imaginative interaction transcending reality, minimalist VR environment, and spatial auditory stimuli enhanced immersion, engagement and emotional connections within the memorial space.

\paragraph{\textbf{Design choice 1: Imaginative Interaction Transcending Reality}}
Imaginative interaction transcending reality were effective in directing Visitors' attention to the Owner’s stories and conveying the content of the memorial space. For example, P7V5 mentioned when he saw life-sized photographs in the photo room. experienced joy, as if he had returned to the past when he raised his children. Participants did not show any discomfort or difficulty when interacting with these imaginative interactions. 
When developing VR memorial spaces, leveraging imaginvative interactions can effectively draw attention to contents’ narratives and enhance users’ connection with the space. To maximize their impact, it is essential to ensure these interactions are both emotionally engaging and seamlessly integrated, so they feel intuitive rather than disruptive to the overall experience.

\paragraph{\textbf{Design choice 2: Minimalist VR Environment}}
A minimal visual design—by unifying walls and floors in light-gray and applying color and texture only to objects with personal significance—helped users explore the VR space naturally without confusion. During the experiment, participants easily identified and interacted with specific objects in the space. For instance, among several objects on a bookshelf, participants immediately moved toward and interacted with a photo album highlighted with color, demonstrating the study's intended focus on "prioritizing key content." 
However, some participants noted that the white space diminished the warmth and emotional comfort they expected in a memorial space, instead creating a cold and desolate atmosphere (P3O2). This suggests that minimal visual design may not work positively for all users. Balancing the visual complexity of VR memorial spaces is crucial, as preferences for simplicity or detail vary. Considering how users perceive and engage with visual elements can ensure the space remains both accessible and emotionally meaningful for diverse Visitors.

\paragraph{\textbf{Design choice 3: Spatial Auditory Stimuli}} Auditory elements in VR memorial spaces can add emotional depth and foster personal connections. For example, a Visitor lingered near a piano while recalling shared memories with the Owner as a familiar piece of music played. This highlights how soundscapes can guide users toward meaningful reflection and create a stronger sense of presence. Auditory elements in VR memorial spaces can evoke deeply personal memories, as each individual may associate different sounds or music with their experiences. While this study utilized proximity-based audio interactions, VR offers the potential to explore more dynamic audio interactions to enhance user's immersive experience \cite{37Asakura2025, 34Liu2025}. Designing these interactions to align with the narrative and emotional context of each space can strengthen users’ immersion and enhance the memorial experience.

\section{CONCLUSION}

This study explores the design and experience of VR memorial space, addressing a gap in HCI and the thanatosensitive design. Using a Research through Design (RtD) approach, we investigated how personalized VR memorial space created with stakeholders can foster posthumous relationships, deepen understanding of the deceased, and inspire new forms of connection. By bridging technology, memory, and human connection, this research advances both theoretical and practical insights into how technology can support emotional and relational needs in the context of loss.

Despite these contributions, our work has two key limitations. First, the participant group was limited in size and diversity, with only eight individuals. While this aligns with RtD’s focus on in-depth exploration over generalizability, future research could involve participants from diverse cultural and relational backgrounds to better understand VR memorial spaces in different contexts. Second, the study focused on the initial implementation and simulated use of these spaces after the Owner's passing, without capturing their long-term impacts on relationships and memorial experiences over time. Future studies could explore how these spaces evolve in meaning, both during the Owner’s lifetime and after their passing, providing deeper insights into their role in sustaining and transforming relationships.

We hope this work inspires future exploration into how VR memorial spaces can serve as mediums for fostering relationships and reflecting on life’s narratives. By emphasizing stakeholder engagement in the design process, this study lays a foundation for developing more inclusive and emotionally resonant VR environments.

\bibliographystyle{ACM-Reference-Format}
\bibliography{8bib}


\begin{thebibliography}{32}

%%% the Web address.

\ifx \showCODEN    \undefined \def \showCODEN     #1{\unskip}     \fi
\ifx \showISBNx    \undefined \def \showISBNx     #1{\unskip}     \fi
\ifx \showISBNxiii \undefined \def \showISBNxiii  #1{\unskip}     \fi
\ifx \showISSN     \undefined \def \showISSN      #1{\unskip}     \fi
\ifx \showLCCN     \undefined \def \showLCCN      #1{\unskip}     \fi
\ifx \shownote     \undefined \def \shownote      #1{#1}          \fi
\ifx \showarticletitle \undefined \def \showarticletitle #1{#1}   \fi
\ifx \showURL      \undefined \def \showURL       {\relax}        \fi
% invisible to TeX
\providecommand\bibfield[2]{#2}
\providecommand\bibinfo[2]{#2}
\providecommand\natexlab[1]{#1}
\providecommand\showeprint[2][]{arXiv:#2}

\bibitem[Albers and Hassenzahl(2024)]%
        {30Albers2024}
\bibfield{author}{\bibinfo{person}{Ruben Albers} {and} \bibinfo{person}{Marc Hassenzahl}.} \bibinfo{year}{2024}\natexlab{}.
\newblock \showarticletitle{Let’s Talk About Death: Existential Conversations with Chatbots}. In \bibinfo{booktitle}{\emph{Proceedings of the 2024 CHI Conference on Human Factors in Computing Systems}} (Honolulu, HI, USA) \emph{(\bibinfo{series}{CHI '24})}. \bibinfo{publisher}{Association for Computing Machinery}, \bibinfo{address}{New York, NY, USA}, Article \bibinfo{articleno}{529}, \bibinfo{numpages}{14}~pages.
\newblock
\showISBNx{9798400703300}
\href{https://doi.org/10.1145/3613904.3642421}{doi:\nolinkurl{10.1145/3613904.3642421}}


\bibitem[Albers et~al\mbox{.}(2023)]%
        {22Albers2023}
\bibfield{author}{\bibinfo{person}{Ruben Albers}, \bibinfo{person}{Shadan Sadeghian}, \bibinfo{person}{Matthias Laschke}, {and} \bibinfo{person}{Marc Hassenzahl}.} \bibinfo{year}{2023}\natexlab{}.
\newblock \showarticletitle{Dying, Death, and the Afterlife in Human-Computer Interaction. A Scoping Review.}. In \bibinfo{booktitle}{\emph{Proceedings of the 2023 CHI Conference on Human Factors in Computing Systems}} (Hamburg, Germany) \emph{(\bibinfo{series}{CHI '23})}. \bibinfo{publisher}{Association for Computing Machinery}, \bibinfo{address}{New York, NY, USA}, Article \bibinfo{articleno}{302}, \bibinfo{numpages}{16}~pages.
\newblock
\showISBNx{9781450394215}
\href{https://doi.org/10.1145/3544548.3581199}{doi:\nolinkurl{10.1145/3544548.3581199}}


\bibitem[Allison et~al\mbox{.}(2024)]%
        {40Allison2024}
\bibfield{author}{\bibinfo{person}{Fraser Allison}, \bibinfo{person}{Bjørn Nansen}, \bibinfo{person}{Martin Gibbs}, \bibinfo{person}{Michael Arnold}, \bibinfo{person}{Samuel Holleran}, {and} \bibinfo{person}{Tamara Kohn}.} \bibinfo{year}{2024}\natexlab{}.
\newblock \showarticletitle{Reimagining memorial spaces through digital technologies: A typology of CemTech}.
\newblock \bibinfo{journal}{\emph{Death Studies}} \bibinfo{volume}{48}, \bibinfo{number}{8} (\bibinfo{year}{2024}), \bibinfo{pages}{790--800}.
\newblock
\href{https://doi.org/10.1080/07481187.2023.2276308}{doi:\nolinkurl{10.1080/07481187.2023.2276308}}
\showeprint{https://doi.org/10.1080/07481187.2023.2276308}
\newblock
\shownote{PMID: 37910087}.


\bibitem[Asakura(2025)]%
        {37Asakura2025}
\bibfield{author}{\bibinfo{person}{Takumi Asakura}.} \bibinfo{year}{2025}\natexlab{}.
\newblock \showarticletitle{Interaction effect of reproducing spatialized human-induced and environmental sounds from nearby sources on the social Simon effect}.
\newblock \bibinfo{journal}{\emph{Applied Acoustics}}  \bibinfo{volume}{228} (\bibinfo{year}{2025}), \bibinfo{pages}{110343}.
\newblock
\showISSN{0003-682X}
\href{https://doi.org/10.1016/j.apacoust.2024.110343}{doi:\nolinkurl{10.1016/j.apacoust.2024.110343}}


\bibitem[Bell(2006)]%
        {4Bell2006}
\bibfield{author}{\bibinfo{person}{Genevieve Bell}.} \bibinfo{year}{2006}\natexlab{}.
\newblock \showarticletitle{No More SMS from Jesus: Ubicomp, Religion and Techno-spiritual Practices}. In \bibinfo{booktitle}{\emph{UbiComp 2006: Ubiquitous Computing}}, \bibfield{editor}{\bibinfo{person}{Paul Dourish} {and} \bibinfo{person}{Adrian Friday}} (Eds.). \bibinfo{publisher}{Springer Berlin Heidelberg}, \bibinfo{address}{Berlin, Heidelberg}, \bibinfo{pages}{141--158}.
\newblock
\showISBNx{978-3-540-39635-2}


\bibitem[Bhide et~al\mbox{.}(2019)]%
        {33Bhide2019}
\bibfield{author}{\bibinfo{person}{Saylee Bhide}, \bibinfo{person}{Elizabeth Goins}, {and} \bibinfo{person}{Joe Geigel}.} \bibinfo{year}{2019}\natexlab{}.
\newblock \showarticletitle{Experimental Analysis of Spatial Sound for Storytelling in Virtual Reality}. In \bibinfo{booktitle}{\emph{Interactive Storytelling: 12th International Conference on Interactive Digital Storytelling, ICIDS 2019, Little Cottonwood Canyon, UT, USA, November 19–22, 2019, Proceedings}} (Little Cottonwood Canyon, UT, USA). \bibinfo{publisher}{Springer-Verlag}, \bibinfo{address}{Berlin, Heidelberg}, \bibinfo{pages}{3–7}.
\newblock
\showISBNx{978-3-030-33893-0}
\href{https://doi.org/10.1007/978-3-030-33894-7_1}{doi:\nolinkurl{10.1007/978-3-030-33894-7_1}}


\bibitem[Chen et~al\mbox{.}(2021)]%
        {16Janet2021}
\bibfield{author}{\bibinfo{person}{Janet~X. Chen}, \bibinfo{person}{Francesco Vitale}, {and} \bibinfo{person}{Joanna McGrenere}.} \bibinfo{year}{2021}\natexlab{}.
\newblock \showarticletitle{What Happens After Death? Using a Design Workbook to Understand User Expectations for Preparing their Data}. In \bibinfo{booktitle}{\emph{Proceedings of the 2021 CHI Conference on Human Factors in Computing Systems}} (Yokohama, Japan) \emph{(\bibinfo{series}{CHI '21})}. \bibinfo{publisher}{Association for Computing Machinery}, \bibinfo{address}{New York, NY, USA}, Article \bibinfo{articleno}{169}, \bibinfo{numpages}{13}~pages.
\newblock
\showISBNx{9781450380966}
\href{https://doi.org/10.1145/3411764.3445359}{doi:\nolinkurl{10.1145/3411764.3445359}}


\bibitem[Facebook(nd)]%
        {1facebook}
\bibfield{author}{\bibinfo{person}{Facebook}.} \bibinfo{year}{[n.d.]}\natexlab{}.
\newblock \bibinfo{title}{About Memorialized Accounts, Facebook Help Center}.
\newblock \bibinfo{howpublished}{\url{https://www.facebook.com/help/1017717331640041/}}.
\newblock
\newblock
\shownote{Accessed: 2025-01-23}.


\bibitem[Gaver(2012)]%
        {15Gaver2012}
\bibfield{author}{\bibinfo{person}{William Gaver}.} \bibinfo{year}{2012}\natexlab{}.
\newblock \showarticletitle{What should we expect from research through design?}. In \bibinfo{booktitle}{\emph{Proceedings of the SIGCHI Conference on Human Factors in Computing Systems}} (Austin, Texas, USA) \emph{(\bibinfo{series}{CHI '12})}. \bibinfo{publisher}{Association for Computing Machinery}, \bibinfo{address}{New York, NY, USA}, \bibinfo{pages}{937–946}.
\newblock
\showISBNx{9781450310154}
\href{https://doi.org/10.1145/2207676.2208538}{doi:\nolinkurl{10.1145/2207676.2208538}}


\bibitem[Gulotta et~al\mbox{.}(2017a)]%
        {7Gulotta2017}
\bibfield{author}{\bibinfo{person}{Rebecca Gulotta}, \bibinfo{person}{Aisling Kelliher}, {and} \bibinfo{person}{Jodi Forlizzi}.} \bibinfo{year}{2017}\natexlab{a}.
\newblock \showarticletitle{Digital Systems and the Experience of Legacy}. In \bibinfo{booktitle}{\emph{Proceedings of the 2017 Conference on Designing Interactive Systems}} (Edinburgh, United Kingdom) \emph{(\bibinfo{series}{DIS '17})}. \bibinfo{publisher}{Association for Computing Machinery}, \bibinfo{address}{New York, NY, USA}, \bibinfo{pages}{663–674}.
\newblock
\showISBNx{9781450349222}
\href{https://doi.org/10.1145/3064663.3064731}{doi:\nolinkurl{10.1145/3064663.3064731}}


\bibitem[Gulotta et~al\mbox{.}(2017b)]%
        {19Gulotta2017}
\bibfield{author}{\bibinfo{person}{Rebecca Gulotta}, \bibinfo{person}{Aisling Kelliher}, {and} \bibinfo{person}{Jodi Forlizzi}.} \bibinfo{year}{2017}\natexlab{b}.
\newblock \showarticletitle{Digital Systems and the Experience of Legacy}. In \bibinfo{booktitle}{\emph{Proceedings of the 2017 Conference on Designing Interactive Systems}} (Edinburgh, United Kingdom) \emph{(\bibinfo{series}{DIS '17})}. \bibinfo{publisher}{Association for Computing Machinery}, \bibinfo{address}{New York, NY, USA}, \bibinfo{pages}{663–674}.
\newblock
\showISBNx{9781450349222}
\href{https://doi.org/10.1145/3064663.3064731}{doi:\nolinkurl{10.1145/3064663.3064731}}


\bibitem[Gulotta et~al\mbox{.}(2013)]%
        {17Gulotta2013}
\bibfield{author}{\bibinfo{person}{Rebecca Gulotta}, \bibinfo{person}{William Odom}, \bibinfo{person}{Jodi Forlizzi}, {and} \bibinfo{person}{Haakon Faste}.} \bibinfo{year}{2013}\natexlab{}.
\newblock \showarticletitle{Digital artifacts as legacy: exploring the lifespan and value of digital data}. In \bibinfo{booktitle}{\emph{Proceedings of the SIGCHI Conference on Human Factors in Computing Systems}} (Paris, France) \emph{(\bibinfo{series}{CHI '13})}. \bibinfo{publisher}{Association for Computing Machinery}, \bibinfo{address}{New York, NY, USA}, \bibinfo{pages}{1813–1822}.
\newblock
\showISBNx{9781450318990}
\href{https://doi.org/10.1145/2470654.2466240}{doi:\nolinkurl{10.1145/2470654.2466240}}


\bibitem[Guo et~al\mbox{.}(2021)]%
        {Guo2021EffectsOV}
\bibfield{author}{\bibinfo{person}{Fu Guo}, \bibinfo{person}{Jiahao Chen}, \bibinfo{person}{Ming ming Li}, \bibinfo{person}{Wei Lyu}, {and} \bibinfo{person}{Junjie Zhang}.} \bibinfo{year}{2021}\natexlab{}.
\newblock \showarticletitle{Effects of visual complexity on user search behavior and satisfaction: an eye-tracking study of mobile news apps}.
\newblock \bibinfo{journal}{\emph{Universal Access in the Information Society}}  \bibinfo{volume}{21} (\bibinfo{year}{2021}), \bibinfo{pages}{795 -- 808}.
\newblock
\urldef\tempurl%
\url{https://api.semanticscholar.org/CorpusID:236572312}
\showURL{%
\tempurl}


\bibitem[H\"{a}kkil\"{a} et~al\mbox{.}(2019)]%
        {11Hakkila2019}
\bibfield{author}{\bibinfo{person}{Jonna H\"{a}kkil\"{a}}, \bibinfo{person}{Petri Hannula}, \bibinfo{person}{Elina Luiro}, \bibinfo{person}{Emilia Launne}, \bibinfo{person}{Sanni Mustonen}, \bibinfo{person}{Toni Westerlund}, {and} \bibinfo{person}{Ashley Colley}.} \bibinfo{year}{2019}\natexlab{}.
\newblock \showarticletitle{Visiting a virtual graveyard: designing virtual reality cultural heritage experiences}. In \bibinfo{booktitle}{\emph{Proceedings of the 18th International Conference on Mobile and Ubiquitous Multimedia}} (Pisa, Italy) \emph{(\bibinfo{series}{MUM '19})}. \bibinfo{publisher}{Association for Computing Machinery}, \bibinfo{address}{New York, NY, USA}, Article \bibinfo{articleno}{56}, \bibinfo{numpages}{4}~pages.
\newblock
\showISBNx{9781450376242}
\href{https://doi.org/10.1145/3365610.3368425}{doi:\nolinkurl{10.1145/3365610.3368425}}


\bibitem[H\"{a}kkil\"{a} et~al\mbox{.}(2024)]%
        {12Hakkila2024}
\bibfield{author}{\bibinfo{person}{Jonna H\"{a}kkil\"{a}}, \bibinfo{person}{Sanni Mustonen}, \bibinfo{person}{Juha-Matti Taikina-Aho}, \bibinfo{person}{Jemina Colley}, \bibinfo{person}{Miika Puljuj\"{a}rvi}, {and} \bibinfo{person}{Siiri Paananen}.} \bibinfo{year}{2024}\natexlab{}.
\newblock \showarticletitle{Designing the Experience for Virtual Graveyard Heritage Sites}. In \bibinfo{booktitle}{\emph{Proceedings of the International Conference on Mobile and Ubiquitous Multimedia}} \emph{(\bibinfo{series}{MUM '24})}. \bibinfo{publisher}{Association for Computing Machinery}, \bibinfo{address}{New York, NY, USA}, \bibinfo{pages}{544–548}.
\newblock
\showISBNx{9798400712838}
\href{https://doi.org/10.1145/3701571.3703377}{doi:\nolinkurl{10.1145/3701571.3703377}}


\bibitem[Husung and Langbehn(2019)]%
        {31Husung2019}
\bibfield{author}{\bibinfo{person}{Malte Husung} {and} \bibinfo{person}{Eike Langbehn}.} \bibinfo{year}{2019}\natexlab{}.
\newblock \showarticletitle{Of Portals and Orbs: An Evaluation of Scene Transition Techniques for Virtual Reality}. In \bibinfo{booktitle}{\emph{Proceedings of Mensch Und Computer 2019}} (Hamburg, Germany) \emph{(\bibinfo{series}{MuC '19})}. \bibinfo{publisher}{Association for Computing Machinery}, \bibinfo{address}{New York, NY, USA}, \bibinfo{pages}{245–254}.
\newblock
\showISBNx{9781450371988}
\href{https://doi.org/10.1145/3340764.3340779}{doi:\nolinkurl{10.1145/3340764.3340779}}


\bibitem[Kim et~al\mbox{.}(2024)]%
        {25Kim2024}
\bibfield{author}{\bibinfo{person}{Jieun Kim}, \bibinfo{person}{Daisuke Uriu}, \bibinfo{person}{Giulia Barbareschi}, \bibinfo{person}{Youichi Kamiyama}, {and} \bibinfo{person}{Kouta Minamizawa}.} \bibinfo{year}{2024}\natexlab{}.
\newblock \showarticletitle{Maintaining Continuing Bonds in Bereavement: A Participatory Design Process of Be.side}. In \bibinfo{booktitle}{\emph{Proceedings of the 2024 CHI Conference on Human Factors in Computing Systems}} (Honolulu, HI, USA) \emph{(\bibinfo{series}{CHI '24})}. \bibinfo{publisher}{Association for Computing Machinery}, \bibinfo{address}{New York, NY, USA}, Article \bibinfo{articleno}{1002}, \bibinfo{numpages}{15}~pages.
\newblock
\showISBNx{9798400703300}
\href{https://doi.org/10.1145/3613904.3642386}{doi:\nolinkurl{10.1145/3613904.3642386}}


\bibitem[Kwon et~al\mbox{.}(2021)]%
        {8Kwon2021}
\bibfield{author}{\bibinfo{person}{Soonho Kwon}, \bibinfo{person}{Eunsol Choi}, \bibinfo{person}{Minseok Kim}, \bibinfo{person}{Sunah Hwang}, \bibinfo{person}{Dongwoo Kim}, {and} \bibinfo{person}{Younah Kang}.} \bibinfo{year}{2021}\natexlab{}.
\newblock \showarticletitle{What Happens to My Instagram Account After I Die? Re-imagining Social Media as a Commemorative Space for Remembrance and Recovery}. In \bibinfo{booktitle}{\emph{Human-Computer Interaction – INTERACT 2021: 18th IFIP TC 13 International Conference, Bari, Italy, August 30 – September 3, 2021, Proceedings, Part II}} (Bari, Italy). \bibinfo{publisher}{Springer-Verlag}, \bibinfo{address}{Berlin, Heidelberg}, \bibinfo{pages}{449–467}.
\newblock
\showISBNx{978-3-030-85615-1}
\href{https://doi.org/10.1007/978-3-030-85616-8_26}{doi:\nolinkurl{10.1007/978-3-030-85616-8_26}}


\bibitem[Liu et~al\mbox{.}(2025)]%
        {34Liu2025}
\bibfield{author}{\bibinfo{person}{Tong Liu}, \bibinfo{person}{Yi Xiao}, \bibinfo{person}{Mingwei Hu}, \bibinfo{person}{Hao Sha}, \bibinfo{person}{Shining Ma}, \bibinfo{person}{Boyu Gao}, \bibinfo{person}{Shihui Guo}, \bibinfo{person}{Yue Liu}, {and} \bibinfo{person}{Weitao Song}.} \bibinfo{year}{2025}\natexlab{}.
\newblock \showarticletitle{AudioGest: Gesture-Based Interaction for Virtual Reality Using Audio Devices}.
\newblock \bibinfo{journal}{\emph{IEEE Transactions on Visualization and Computer Graphics}} \bibinfo{volume}{31}, \bibinfo{number}{2} (\bibinfo{year}{2025}), \bibinfo{pages}{1569--1581}.
\newblock
\href{https://doi.org/10.1109/TVCG.2024.3397868}{doi:\nolinkurl{10.1109/TVCG.2024.3397868}}


\bibitem[Locasto et~al\mbox{.}(2011)]%
        {18Locasto2011}
\bibfield{author}{\bibinfo{person}{Michael~E. Locasto}, \bibinfo{person}{Michael Massimi}, {and} \bibinfo{person}{Peter~J. DePasquale}.} \bibinfo{year}{2011}\natexlab{}.
\newblock \showarticletitle{Security and privacy considerations in digital death}. In \bibinfo{booktitle}{\emph{Proceedings of the 2011 New Security Paradigms Workshop}} (Marin County, California, USA) \emph{(\bibinfo{series}{NSPW '11})}. \bibinfo{publisher}{Association for Computing Machinery}, \bibinfo{address}{New York, NY, USA}, \bibinfo{pages}{1–10}.
\newblock
\showISBNx{9781450310789}
\href{https://doi.org/10.1145/2073276.2073278}{doi:\nolinkurl{10.1145/2073276.2073278}}


\bibitem[Lopes et~al\mbox{.}(2014)]%
        {39Lopes2014}
\bibfield{author}{\bibinfo{person}{Aron~Daniel Lopes}, \bibinfo{person}{Cristiano Maciel}, {and} \bibinfo{person}{Vinicius~Carvalho Pereira}.} \bibinfo{year}{2014}\natexlab{}.
\newblock \showarticletitle{Virtual homage to the dead: an analysis of digital memorials in the social web}. In \bibinfo{booktitle}{\emph{International Conference on Social Computing and Social Media}}. Springer, \bibinfo{pages}{67--78}.
\newblock


\bibitem[Maddrell(2013)]%
        {21Maddrell2013}
\bibfield{author}{\bibinfo{person}{Avril Maddrell}.} \bibinfo{year}{2013}\natexlab{}.
\newblock \showarticletitle{Living with the Deceased: Absence, Presence and Absence-Presence}.
\newblock \bibinfo{journal}{\emph{cultural geographies}}  \bibinfo{volume}{20} (\bibinfo{date}{09} \bibinfo{year}{2013}), \bibinfo{pages}{501--522}.
\newblock
\href{https://doi.org/10.1177/1474474013482806}{doi:\nolinkurl{10.1177/1474474013482806}}


\bibitem[Massimi and Charise(2009)]%
        {6Massimi2009}
\bibfield{author}{\bibinfo{person}{Michael Massimi} {and} \bibinfo{person}{Andrea Charise}.} \bibinfo{year}{2009}\natexlab{}.
\newblock \showarticletitle{Dying, death, and mortality: towards thanatosensitivity in HCI}. In \bibinfo{booktitle}{\emph{CHI '09 Extended Abstracts on Human Factors in Computing Systems}} (Boston, MA, USA) \emph{(\bibinfo{series}{CHI EA '09})}. \bibinfo{publisher}{Association for Computing Machinery}, \bibinfo{address}{New York, NY, USA}, \bibinfo{pages}{2459–2468}.
\newblock
\showISBNx{9781605582474}
\href{https://doi.org/10.1145/1520340.1520349}{doi:\nolinkurl{10.1145/1520340.1520349}}


\bibitem[Moncur et~al\mbox{.}(2015)]%
        {27Wendy2015}
\bibfield{author}{\bibinfo{person}{Wendy Moncur}, \bibinfo{person}{Miriam Julius}, \bibinfo{person}{Elise Van Den~Hoven}, {and} \bibinfo{person}{David Kirk}.} \bibinfo{year}{2015}\natexlab{}.
\newblock \showarticletitle{Story Shell: the participatory design of a bespoke digital memorial}. In \bibinfo{booktitle}{\emph{Proceedings of 4th Participatory Innovation Conference}}. \bibinfo{pages}{470--477}.
\newblock


\bibitem[Moncur and Kirk(2014)]%
        {3Moncur2014}
\bibfield{author}{\bibinfo{person}{Wendy Moncur} {and} \bibinfo{person}{David Kirk}.} \bibinfo{year}{2014}\natexlab{}.
\newblock \showarticletitle{An emergent framework for digital memorials}. In \bibinfo{booktitle}{\emph{Proceedings of the 2014 Conference on Designing Interactive Systems}} (Vancouver, BC, Canada) \emph{(\bibinfo{series}{DIS '14})}. \bibinfo{publisher}{Association for Computing Machinery}, \bibinfo{address}{New York, NY, USA}, \bibinfo{pages}{965–974}.
\newblock
\showISBNx{9781450329026}
\href{https://doi.org/10.1145/2598510.2598516}{doi:\nolinkurl{10.1145/2598510.2598516}}


\bibitem[Petersson and Wingren(2011)]%
        {20Petersson2011}
\bibfield{author}{\bibinfo{person}{Anna Petersson} {and} \bibinfo{person}{Carola Wingren}.} \bibinfo{year}{2011}\natexlab{}.
\newblock \showarticletitle{Designing a memorial place: Continuing care, passage landscapes and future memories}.
\newblock \bibinfo{journal}{\emph{Mortality}}  \bibinfo{volume}{16} (\bibinfo{date}{02} \bibinfo{year}{2011}), \bibinfo{pages}{54--69}.
\newblock
\href{https://doi.org/10.1080/13576275.2011.536369}{doi:\nolinkurl{10.1080/13576275.2011.536369}}


\bibitem[Pixar(2017)]%
        {Coco2017}
\bibfield{author}{\bibinfo{person}{Disney Pixar}.} \bibinfo{year}{2017}\natexlab{}.
\newblock \bibinfo{title}{Coco}.
\newblock \bibinfo{howpublished}{Film, Directed by Lee Unkrich and Adrian Molina. Walt Disney Pictures}.
\newblock


\bibitem[She et~al\mbox{.}(2021)]%
        {36She2021}
\bibfield{author}{\bibinfo{person}{Wan-Jou She}, \bibinfo{person}{Panote Siriaraya}, \bibinfo{person}{Chee~Siang Ang}, {and} \bibinfo{person}{Holly~Gwen Prigerson}.} \bibinfo{year}{2021}\natexlab{}.
\newblock \showarticletitle{Living Memory Home: Understanding Continuing Bond in the Digital Age through Backstage Grieving}. In \bibinfo{booktitle}{\emph{Proceedings of the 2021 CHI Conference on Human Factors in Computing Systems}} (Yokohama, Japan) \emph{(\bibinfo{series}{CHI '21})}. \bibinfo{publisher}{Association for Computing Machinery}, \bibinfo{address}{New York, NY, USA}, Article \bibinfo{articleno}{546}, \bibinfo{numpages}{14}~pages.
\newblock
\showISBNx{9781450380966}
\href{https://doi.org/10.1145/3411764.3445336}{doi:\nolinkurl{10.1145/3411764.3445336}}


\bibitem[Simonsen and Robertson(2013)]%
        {38Simonsen2012}
\bibfield{author}{\bibinfo{person}{Jesper Simonsen} {and} \bibinfo{person}{Toni Robertson}.} \bibinfo{year}{2013}\natexlab{}.
\newblock \bibinfo{booktitle}{\emph{Routledge international handbook of participatory design}}. Vol.~\bibinfo{volume}{711}.
\newblock \bibinfo{publisher}{Routledge New York}.
\newblock


\bibitem[Uriu et~al\mbox{.}(2021)]%
        {13Uriu2021}
\bibfield{author}{\bibinfo{person}{Daisuke Uriu}, \bibinfo{person}{Noriyasu Obushi}, \bibinfo{person}{Zendai Kashino}, \bibinfo{person}{Atsushi Hiyama}, {and} \bibinfo{person}{Masahiko Inami}.} \bibinfo{year}{2021}\natexlab{}.
\newblock \showarticletitle{Floral Tribute Ritual in Virtual Reality: Design and Validation of SenseVase with Virtual Memorial}. In \bibinfo{booktitle}{\emph{Proceedings of the 2021 CHI Conference on Human Factors in Computing Systems}} (Yokohama, Japan) \emph{(\bibinfo{series}{CHI '21})}. \bibinfo{publisher}{Association for Computing Machinery}, \bibinfo{address}{New York, NY, USA}, Article \bibinfo{articleno}{628}, \bibinfo{numpages}{15}~pages.
\newblock
\showISBNx{9781450380966}
\href{https://doi.org/10.1145/3411764.3445216}{doi:\nolinkurl{10.1145/3411764.3445216}}


\bibitem[Wyche et~al\mbox{.}(2006)]%
        {5Wyche2006}
\bibfield{author}{\bibinfo{person}{Susan~P. Wyche}, \bibinfo{person}{Gillian~R. Hayes}, \bibinfo{person}{Lonnie~D. Harvel}, {and} \bibinfo{person}{Rebecca~E. Grinter}.} \bibinfo{year}{2006}\natexlab{}.
\newblock \showarticletitle{Technology in spiritual formation: an exploratory study of computer mediated religious communications}. In \bibinfo{booktitle}{\emph{Proceedings of the 2006 20th Anniversary Conference on Computer Supported Cooperative Work}} (Banff, Alberta, Canada) \emph{(\bibinfo{series}{CSCW '06})}. \bibinfo{publisher}{Association for Computing Machinery}, \bibinfo{address}{New York, NY, USA}, \bibinfo{pages}{199–208}.
\newblock
\showISBNx{1595932496}
\href{https://doi.org/10.1145/1180875.1180908}{doi:\nolinkurl{10.1145/1180875.1180908}}


\bibitem[Zimmerman et~al\mbox{.}(2007)]%
        {14Zimmerman2007}
\bibfield{author}{\bibinfo{person}{John Zimmerman}, \bibinfo{person}{Jodi Forlizzi}, {and} \bibinfo{person}{Shelley Evenson}.} \bibinfo{year}{2007}\natexlab{}.
\newblock \showarticletitle{Research through design as a method for interaction design research in HCI}. In \bibinfo{booktitle}{\emph{Proceedings of the SIGCHI Conference on Human Factors in Computing Systems}} (San Jose, California, USA) \emph{(\bibinfo{series}{CHI '07})}. \bibinfo{publisher}{Association for Computing Machinery}, \bibinfo{address}{New York, NY, USA}, \bibinfo{pages}{493–502}.
\newblock
\showISBNx{9781595935939}
\href{https://doi.org/10.1145/1240624.1240704}{doi:\nolinkurl{10.1145/1240624.1240704}}


\end{thebibliography}

\end{document}